\documentclass[12pt]{article} 
\usepackage[dvips]{graphicx}
\begin{document}
\title{Variational approach for the quantum Zakharov system}
\author{F. Haas\footnote{ferhaas@exatas.unisinos.br}\\
Universidade do Vale do Rio dos Sinos - UNISINOS \\
Unidade de Exatas e Tecnol\'ogicas \\
Av. Unisinos, 950\\
93022--000 S\~ao Leopoldo, RS, Brazil}
\maketitle
\begin{abstract}
The quantum Zakharov system is described in terms of a Lagrangian formalism. A time-dependent Gaussian trial function approach for the envelope electric field and the low-frequency part of the density fluctuation leads to a coupled, nonlinear system of ordinary differential equations. In the semiclassic case, linear stability analysis of this dynamical system shows a destabilizing r\^ole played by quantum effects. Arbitrary value of the quantum effects are also considered, yielding the ultimate destruction of the localized, Gaussian trial solution. Numerical simulations are shown both for the semiclassic and the full quantum cases.    
\end{abstract}
\maketitle

\section{Introduction}
Quantum plasmas have received much attention in recent times, especially because of the ongoing miniaturization of ultrasmall electronic devices and micromechanical systems \cite{Markowich} and to the relevance of quantum effects for intense laser-plasmas  \cite{Rascol} and for dense astrophysical objects \cite{Opher}. Frequently, the de Broglie wavelength of the charge carriers (electrons, positrons, holes) of these systems is comparable to the characteristic dimensions of the system, making a quantum treatment unavoidable. Advances in the area includes construction of quantum ion-acoustic waves \cite{H}, quantum magnetohydrodynamics theories \cite{HH}, quantum beam instabilities \cite{Haas}--\cite{Haas2} and shear Alfv\'en modes in ultra-cold quantum magnetoplasmas \cite{Shukla2}. New quantum collective excitations have also been identified for ultra-cold dusty plasmas \cite{Stenflo}--\cite{Misra}, where quantum effects can be used for plasma diagnostics. Recently, spin effects have been included for non relativistic quantum plasmas \cite{mm, mmm}. Possible decisive  applications of spin effects in quantum plasmas can appear in solid state plasmas as well as in pulsars and magnetars, with very intense magnetic fields (greater than $10^8 T$). A more detailed review on quantum plasma models and their range of validity can be found in \cite{Manfred}.

The quantum Zakharov equations \cite{Garcia}, the subject of the present work, form a set of coupled, nonlinear partial differential equations for the envelope electric field and the slow part of the density fluctuation in an electron-ion electrostatic quantum plasma. It models the interaction between quantum Langmuir and quantum ion-acoustic waves. Exactly as for the classical Zakharov system \cite{Zakharov}, the derivation of the quantum Zakharov equations comes from the existence of two time scales, a fast one associated to quantum Langmuir waves, and a slow one associated to quantum ion-acoustic waves. Sample applications can be found for quantum decay and four-wave instabilities, with relevant changes of the  classical dispersions \cite{Garcia}. The quantum Zakharov system was also analyzed for the  enhancement of modulational instabilities due to combination of partial coherence and quantum corrections \cite{Marklund}. More recently, the coupling between nonlinear Langmuir waves and electron holes in Wigner-Poisson quantum plasmas was studied via a two time scales formalism \cite{Jovanovic}.

The existence of coherent structures, as soliton solutions for instance, is a relevant issue for any system of evolution equations. As an example from quantum plasmas, stable vortices and dark solitons  have been constructed for Schr\"odinger-Poisson quantum electron plasmas \cite{Shukla}. At quantum scales, the transport of information in ultracold micromechanical systems can be addressed by means of such nonlinear structures. The basic objective of the present work is the investigation of the quantum effects for the existence of localized solutions for the quantum Zakharov system. Unlike the approach of Yang {\it et al.}, where exact bright solitons, gray solitons, W-solitons and M-solitons were found for the quantum Zakharov system \cite{Yang}, here approximate solutions are obtained through a variational formulation and a trial function method. Exact solutions are of course  very relevant, but variational solutions provides more insight on the r\^ole of quantum effects. For instance, the classical Zakharov equations admit the Langmuir soliton solution \cite{Thornhill}. Using a Gaussian ansatz as a trial function extremizing an action functional, one can get information about the perturbation of the Langmuir soliton by quantum effects. A priori, one can expect that wave-packet spreading and tunneling tends to enlarge the width of localized wave solutions. Other possibility is the appearance of instabilities of pure quantum nature, eventually destroying any coherent structure. Besides these considerations, the construction of a variational formulation for the quantum Zakharov equations is important by itself. Notice that the internal vibrations of solitary waves for the classical Zakharov system were analyzed by a variational approach using Gaussian trial functions \cite{Malomed}. The present contribution extends this work to the quantum realm. Similar time-dependent variational methods were also used, for instance, for the nonlinear pulse propagation in optical fibers \cite{Anderson} and for Bose-Einstein condensates \cite{bec}. 

Variational methods can indicate a general tendency of s system for which no general closed form solution is available. For instance, one can study the changes of localized or solitonic trial functions under the changes of a control parameter. The quantum Zakharov equations (see Section II) possess a single dimensionless quantity $H$ measuring the importance of quantum perturbations. It is one basic task of this work, to analyze the changes in Gaussian trial function solutions for the quantum Zakharov system induced by modifications in $H$.  

This work is organized as follows. In section II, the quantum Zakharov system is described by a variational formulation. A variational solution in the form of a Gaussian ansatz is then proposed, in order to reproduce the main properties of the Langmuir soliton solution admitted in the classical limit. This time-dependent trial function approach leads to a dynamical system which can be analyzed for several parameter regimes. In section III, only the first-order quantum correction is retained, yielding a set of two coupled, nonlinear second-order ordinary differential equations for the widths of the envelope electric field and density perturbation. This nonlinear system is analyzed for its linear stability properties as well as for the existence of bounded solutions. In section IV, arbitrary strength of the quantum effects is allowed, resulting in a full system of equations. Further, the ultra-quantum case where quantum effects are the more relevant influence is analyzed, showing the ultimate destruction of the Langmuir soliton due to wave-packet spreading and tunneling. Section V is reserved to the conclusions.  

\section{Variational formulation}
The one-dimensional quantum Zakharov equations reads \cite{Garcia} 
\begin{eqnarray}
\label{ee1}
&&i{\partial E\over\partial t}+{\partial ^2E\over\partial
x^2}-H^2
{\partial ^4E\over\partial x^4}=n\,E\,,\\
\label{ee2}
&&{\partial ^2n\over\partial t^2}-{\partial ^2 n\over\partial
x^2}
+H^2{\partial ^4n\over\partial x^4}={\partial
^2|E|^2\over\partial x^2}\,,
\end{eqnarray}
where $E = E(x,t)$ is the envelope electric field and
$n = n(x,t)$ is the density fluctuation. All quantities are expressed in a convenient dimensionless form. Further, 
\begin{equation}
\label{ee3}
H = {\hbar\, \omega_{i}\over\kappa_B\,T_e} 
\end{equation}
is a parameter expressing the ratio between the ion plasmon energy and the electron thermal energy, where $\hbar$ is the scaled Planck constant, $\kappa_B$ the Boltzmann constant, $\omega_i$ the ion plasma frequency and $T_e$ the electron temperature. The formal classical limit is obtained for $H \equiv 0$, yielding the original Zakharov system. For more details on the derivation of the system (\ref{ee1}--\ref{ee2}) as well as for sample applications, see \cite{Garcia}. 

The one-dimensional quantum Zakharov equations are derived from the  Lagrangian density 
\begin{eqnarray}
{\cal{L}} &=& \frac{i}{2}\left(E^{*}\frac{\partial E}{\partial t} - E\frac{\partial E^{*}}{\partial t}\right) - \left|\frac{\partial E}{\partial x}\right|^2 - \frac{\partial u}{\partial x} \, |E|^2 + \frac{1}{2}\left(\frac{\partial u}{\partial t}\right)^2 - \frac{1}{2}\left(\frac{\partial u}{\partial x}\right)^2 \nonumber \\
\label{e1} &-& H^2 \left|\frac{\partial^2 E}{\partial x^2}\right|^2 - \frac{H^2}{2}\left(\frac{\partial^2 u}{\partial x^2}\right)^2 \,,
\end{eqnarray}
where it was introduced the auxiliary variable $u$ so that 
\begin{equation}
\label{e2}
n = \frac{\partial u}{\partial x} \,.
\end{equation}
Indeed, 
\begin{eqnarray}
\frac{\delta{\cal L}}{\delta E} &=& 0 \Rightarrow - i{\partial E^{*}\over\partial t}+{\partial ^2E^{*}\over\partial
x^2}-H^2
{\partial ^4E^{*}\over\partial x^4}=\frac{\partial u}{\partial x}\,E^{*}\,, \\
\frac{\delta{\cal L}}{\delta E^{*}} &=& 0 \Rightarrow i{\partial E\over\partial t}+{\partial ^2E\over\partial
x^2}-H^2
{\partial ^4E\over\partial x^4}=\frac{\partial u}{\partial x}\,E\,,\\
\frac{\delta{\cal L}}{\delta u} &=& 0 \Rightarrow - \frac{\partial}{\partial x}(|E|^2 + \frac{\partial u}{\partial x}) + \frac{\partial^2 u}{\partial t^2} + H^2 \frac{\partial^4 u}{\partial x^4} = 0 \,.
\end{eqnarray}
The last equation reproduces (\ref{ee2}) after differentiation with respect to $x$. 

The classical Zakharov system is not integrable. However, it admits \cite{Thornhill} the exact Langmuir soliton solution 
\begin{eqnarray}
\label{e4}
E &=& E_0 \sec{\rm\!h}\left(\frac{E_0 x}{\sqrt{2}}\right) \exp\left(\frac{iE_{0}^{2}t}{2}\right) \,,\\
\label{e5}
n &=& - E_{0}^2 \sec{\rm\!h}^{2}\left(\frac{E_0 x}{\sqrt{2}}\right) \,,
\end{eqnarray}
where $E_0$ is an arbitrary real parameter. Strictly, collisions of Langmuir ``solitons" does not simply imply phase shifts between them \cite{Thornhill}, as expected for solitonic objects. Nevertheless, it is the interplay between nonlinear and dispersive terms in the classical Zakharov system which allows for the existence of such coherent structures. Physically, the Langmuir soliton represents a hole in the low frequency part of the electron-ion density maintained self-consistently by the ponderomotive force. It would be desirable to achieve a better understanding of the quantum effects for this soliton solution. A priori one can expect tunneling of electrons trapped in the self-consistent potential, perturbing the remarkable stability of the Langmuir solitons in the classical case. Indeed, isolated classical Langmuir solitons do not decay \cite{Thornhill}.

A variational solution which reproduces the gross features of (\ref{e4}-\ref{e5}) and is at the same time analytically accessible is the Gaussian ansatz
\begin{eqnarray}
\label{e6}
E &=& A \exp(-\frac{x^2}{2a^2} + i\phi + i\kappa x^2) \,,\\
\label{e7}
n &=& - B \exp(-\frac{x^2}{b^2}) \,,
\end{eqnarray}
where $A, B, a, b, \phi$ and $\kappa$ are functions of time. We assume $A$ and $B$ positive to maintain resemblance with (\ref{e4}-\ref{e5}). As for any time-dependent variational method, notice that a main drawback of the Gaussian ansatz is that it does not allow for changes in the shape of the solution. The classical Zakharov system can be treated by a variational approach using a combination of Jacobi elliptic functions \cite{Sharma}, but we use Gaussian functions for the sake of simplicity.   

In order to calculate the Lagrangian $L = \int_{-\infty}^{\infty} {\cal L} dx$ corresponding to (\ref{e6}-\ref{e7}) there is the need of the derivatives $\partial u/\partial x$ and $\partial u/\partial t$ of the auxiliary function $u$. Combining (\ref{e2}) and (\ref{e7}) and introducing $M = Bb$, it follows that 
\begin{equation}
\label{e8}
\frac{\partial u}{\partial t} = \frac{M\dot{b}x}{b^2}\exp(-\frac{x^2}{b^2}) - \frac{\dot{M}}{b}\int_{0}^{x} \exp(-\frac{y^2}{b^2}) dy \,.   
\end{equation}
Inserting the last expression into the integral for the Lagrangian one concludes that it converges if and only if $\dot{M} = 0$, implying a restriction on the allowable variational density functions. Indeed, if $\dot{M} \neq 0$, then the proposed Gaussian ansatz leads to divergence of the Lagrangian.  

$M$ being invariant is consistent with the conservation of the low frequency part of the mass
\begin{equation}
\label{e9}
\frac{1}{\sqrt{\pi}}\int_{-\infty}^{+\infty} n dx = Bb \,,
\end{equation}
the last equality following from the variational solution and with the factor $1/\sqrt{\pi}$ being introduced for convenience. In addition, we note the existence of the conservation of the number of high frequency quanta
\begin{equation}
\label{e10}
N = \frac{1}{\sqrt{\pi}}\int_{-\infty}^{+\infty} |E|^2 dx = A^2 a
\end{equation}
and of the Hamiltonian  
\begin{eqnarray}
{\cal{H}} &=& \frac{1}{\sqrt{\pi}}\int_{-\infty}^{\infty} {[} \, \left|\frac{\partial E}{\partial x}\right|^2 + \frac{\partial u}{\partial x} \, |E|^2 + \frac{1}{2}\left(\frac{\partial u}{\partial t}\right)^2 + \frac{1}{2}\left(\frac{\partial u}{\partial x}\right)^2 \nonumber \\  &+& H^2 \left|\frac{\partial^2 E}{\partial x^2}\right|^2 + \frac{H^2}{2}\left(\frac{\partial^2 u}{\partial x^2}\right)^2 \, {]} \, dx  \nonumber  \\  \label{e3} &=& \frac{N}{2}\left(\frac{1}{a^2} + 4 \kappa^2 a^2 \right) - \frac{M N}{\sqrt{a^2 + b^2}}  + \frac{M^2}{2\sqrt{2}b} + \frac{M^2 \dot{b}^2}{8\sqrt{2}b} \\ &+& \frac{3H^2 N}{4} \left(\frac{1}{a^2} + 4 \kappa^2 a^2 \right)^2 + \frac{H^2 M^2}{2\sqrt{2}b^3} \,. \nonumber 
\end{eqnarray}
The numerical factor $1/\sqrt{\pi}$ was used for convenience again, in the definitions of $N$ and ${\cal H}$, while the last equalities at  (\ref{e10}-\ref{e3}) follows from the proposed variational solution. The quantum Zakharov equations also preserve a momentum functional, but this information is useless in the remaining.

Now we get the effective Lagrangian 
\begin{eqnarray}
L = \sqrt{\pi} \,\, [- A^2 a\dot \phi &-& \frac{A^2 a}{2}\left(a^2 \,\dot\kappa + \frac{1}{a^2} + 4\kappa^2 a^2\right) + \frac{M A^2 a}{\sqrt{a^2 + b^2}}  - \frac{M^2}{2\sqrt{2}b}  \nonumber \\ &+& \frac{M^2 \dot{b}^2}{8\sqrt{2}b}  \label{e11} - \frac{3H^2 A^2 a}{4} \left(\frac{1}{a^2} + 4 \kappa^2 a^2 \right)^2 - \frac{H^2 M^2}{2\sqrt{2}\,b^3}] \,, 
\end{eqnarray}
depending on the dynamical variables  $\phi, \kappa, A, a, b$ and their derivatives. The last two terms at (\ref{e11}) contains the quantum corrections. 

Variation in $\phi$ gives $\dot N = 0$, just reproducing the conservation of the high frequency quanta. Variation in $\kappa$, the so-called chirp function, gives
\begin{equation}
\label{e12} a\dot{a} = 4\kappa a^2 + 12 H^2 \kappa (1 + 4\kappa^2 a^4) \,.
\end{equation}
In contrast to the classical case where the chirp function is easily derived from $a(t)$, in the quantum case $\kappa$ is given in terms of $a$ as the solution of the third-degree equation (\ref{e12}). Hence, expressing $\kappa$ in terms of $a$ and $\dot{a}$ would give equations too cumbersome to be of any value. A reasonable alternative in the semiclassic limit is to solve (\ref{e12}) for $\kappa$ as a power series in $H^2$. Other possibility is to regard (\ref{e12}) as a dynamical equation to be numerically solved for $a(t)$. Both approaches will be considered in what follows. 

Combining independent variations in $a$ and in $A$ gives 
\begin{equation}
\label{e13}
a\dot{\kappa} = \frac{1}{a^3} - 4\kappa^{2}a - \frac{M a}{(a^2 + s^4)^{3/2}} + \frac{3H^2}{a^5} - 48 H^2 \kappa^4 a^3 \,, 
\end{equation}
while varying $b$ gives
\begin{equation}
\label{e14}
\ddot{s} = \frac{1}{s^3} - \frac{2\sqrt{2} \,N s^3}{M (a^2 + s^4)^{3/2}} + \frac{3H^2}{s^7} \,,
\end{equation}
where $b = s^2$.

Equations (\ref{e12}-\ref{e14}) form a complete system for the dynamical variables $a, \kappa$ and $s$ and are the basis for the conclusions in the following. The system (\ref{e12}-\ref{e14}) will be studied in the semiclassic limit, both for linear and nonlinear oscillations, and for arbitrary values of the quantum parameter $H$.

\section{Semiclassic oscillations}
It is of considerable interest to investigate the modifications in the Langmuir soliton induced by small quantum effects. When $H$ is a small parameter, (\ref{e12}) can be solved approximately in powers of $H^2$ yielding 
\begin{equation}
\label{e15}
\kappa = \frac{\dot a}{4a} - H^2 \left(\frac{3\dot a}{4 a^3} + \frac{3\dot{a}^3}{16 a}\right) \,, 
\end{equation}
disregarding higher-order corrections. Then one is left with a coupled, nonlinear system of second-order equations for $a$ and $s$, namely, (\ref{e14}) and
\begin{equation}
\label{e16}
\ddot a = \frac{4}{a^3} - \frac{4 M a}{(a^2 + s^4)^{3/2}} + H^2 \left(\frac{24}{a^5} + \frac{6\dot{a}^2}{a^3} - \frac{12 M}{a (a^2 + s^4)^{3/2}} - \frac{9 M a \dot{a}^2}{(a^2 + s^4)^{3/2}} \right) \,.
\end{equation}
Consistently, the energy (\ref{e3}) evaluated using (\ref{e15}), 
\begin{eqnarray}
\label{e17}
{\cal H} &=& \frac{1}{4}\left(N (\frac{\dot{a}^2}{2} + \frac{2}{a^2}) + \sqrt{2} M^2 (\dot{s}^2 + \frac{1}{s^2}) - \frac{4 M N}{\sqrt{a^2 + s^4}}\right) \\ &+& \frac{H^2}{64}\left(\frac{48 N}{a^4} - \frac{24 N \dot{a}^2}{a^2} - 9 N \dot{a}^4 + \frac{16\sqrt{2}M^2}{s^6}\right) \, \nonumber
\end{eqnarray}
is approximately constant, $d{\cal H}/dt = O(H^4)$ along trajectories of (\ref{e14})-(\ref{e16}), and can be used to check the accuracy of numerical schemes.   

It is relevant to check (\ref{e14})-(\ref{e16}) for the linear stability of fixed points. For the Langmuir soliton (\ref{e4}-\ref{e5}) one have equal values for $M$ and $N$. Since we are mainly interested in the r\^ole of quantum effects for the Langmuir soliton, in the rest of the section we set $M = N$. Since quantum effects are small, one can search for fixed points for the dynamical system as a power series in $H^2$. An easy calculation then yields critical points at $(a,s) = (a_{0},s_{0})$, with 
\begin{eqnarray}
\label{c1}
a_0 &=& \frac{2\sqrt{2}}{M} + \frac{3 H^2 M}{\sqrt{2}} \,,\\
\label{c2}
s_0 &=& \left(\frac{2\sqrt{2}}{M}\right)^{1/2} + \frac{H^2 M^{3/2}}{2^{1/4}} \,,
\end{eqnarray}
disregarding $O(H^4)$ terms. In terms of the original variables $a$ and $b$ these fixed points corresponds to Gaussians of same width at the formal classical limit. Quantum corrections, however, introduce a disturbance: the width $a$ of the Gaussian for the envelope electric field increases less than the width $b$ associated to the density. Moreover, both characteristic lengths increase, pointing for a wave-packet spreading effect.

Considering small deviations $\sim \exp(i\omega t)$ from the equilibrium point, one obtain
\begin{eqnarray}
\label{e18}
\omega^{2} &=& \frac{M^2}{128} \left[24 + 10 M^2 - H^2 M^2 (18 M^2 + 33)\right] \\ &\pm& \frac{M^2}{64} \left[144 + 24 M^2 + 25 M^4  - H^2 M^2 (396 + 177 M^2 + 90 M^4)\right]^{1/2}
 \,.  \nonumber 
\end{eqnarray}
For consistency one could also expand the square root at (\ref{e18}) up to $O(H^2)$ terms, but this would result in a more cumbersome expression. 

Unstable linear oscillations corresponds to solutions with $Im(\omega) < 0$. A straightforward algebra shows that such instabilities are impossible in the formal classical limit $H \equiv 0$. In the quantum case, however, a careful analysis of (\ref{e18}) shows that instabilities are possible when
\begin{equation}
\label{e19}
H^2 > f(M^2) \equiv \frac{144 + 24 M^2 + 25 M^4}{M^2 (396 + 177 M^2 + 90 M^4)} \,.
\end{equation}
This instability condition can be satisfied for small values of $H$. For instance, for $M > 0.6$, $f(M^2) < 1$ at (\ref{e19}). Further increasing $M$ allows for smaller values of $H$. For $M = 5$, one has $H > 1/10$ for instability, an inequality which can be satisfied within the present semiclassic context. In terms of the Langmuir soliton (\ref{e4}-\ref{e5}), $M = 2\sqrt{2}E_0$, showing that large amplitude solitons are more influenced by quantum instabilities.  Figure 1 shows the curve $H^2 = f(M^2)$ separating stable-unstable regions. 

\begin{figure}
\includegraphics{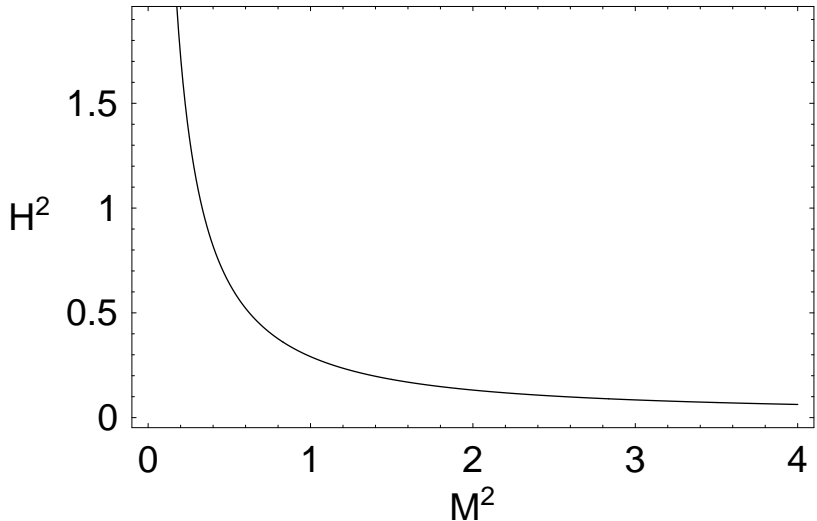}
\caption{\label{fig:epsart1} The curve $H^2 = f(M^2)$ for the instability condition (\ref{e19}). Instability occurs for $H^2 > f(M^2)$}
\end{figure}

The existence of quantum instabilities not present for the classical system is a remarkable fact. It shows that the classical localized solution eventually is smeared out, since the width of the Gaussian is continuously increasing in time. This is a signature of wave-packet spreading and tunneling, because the classical Langmuir soliton is produced by particle trapping in the self-consistent electrostatic potential. Notice, however, that nonlinear effects can suppress the linear quantum instability. 

Unlike the classical case \cite{Malomed}, the system (\ref{e14})-(\ref{e16}) seems to be not described by a pseudo-potential function. This is due to the velocity dependence on the dynamical equation for $a(t)$. Nevertheless, using the energy integral ${\cal H}$ one can get a rough estimate for an escape velocity. In the limit situation where the particle escapes, one has $\dot a = \dot{s} = 0$ when $a \rightarrow \infty$ and $s \rightarrow \infty$. From (\ref{e17}) it gives ${\cal H} = 0$. Now supposing an initial condition at the fixed point (\ref{c1}-\ref{c2}) and taking $\dot{a}(0) = \dot{a}_0 = 0$ for simplicity, one finds a escape velocity $\dot{s}(0) = \dot{s}_0$ such that (for $M = N$)
\begin{equation}
\label{t}
\dot{s}_{0}^2 = \frac{M}{4\sqrt{2}} - \frac{7 H^2 M^3}{64\sqrt{2}} \,.
\end{equation}
Once again, quantum effects act in a tunneling-like manner. Indeed, (\ref{t}) shows that a sufficiently large value of $H$ can produce escaping of the particle, no matter the value of the initial velocity $\dot{s}_0$. The limiting value $H = (4/\sqrt{7})M^{-1}$ for which $\dot{s}_0 = 0$ can be achieved even for the semiclassic case for sufficiently high $M$. For instance, when $M = 10$, the particle eventually escapes for $H > 0.15$, a moderate value. A nonzero value of $\dot{a}_0$ tends to produce even smaller values of $H$ for escaping. 

Figures 2, 3 and 4 shows typical oscillations for the dynamical system (\ref{e14})-(\ref{e16}), with $M = N = 3$, $H = 0.3$. The initial condition is at the fixed point and $\dot{a}(0) = 0$. Also, $\dot{s}_0 = 0.62$. For such parameters, simulations shows unbounded motion for $\dot{s}_0 = 0.64$, which is much less than the classical escape velocity, $\dot{s}_0 = 0.73$, and in good agreement with the critical value $0.59$ arising from the crude estimate (\ref{t}). The Hamiltonian ${\cal H}$ remains  approximately constant at the value $- 0.10$ along the run. Observe the different time-scales for $a(t)$ and $s(t)$. Taking a smaller value of $\dot{s}_0$ gives a more regular, quasi-periodic oscillation pattern, similar to the classical oscillations \cite{Malomed}. However, notice that quantum effects leads to complicated trajectories even for equal values of the invariants $M$ and $N$, approaching the critical value of $\dot{s}_0$ for unbounded motion (see figure 4). Several other runs shows that increasing the value of $\dot{s}_0$ increases the period and the amplitude of the oscillations. Direct comparison between the present simulations and those of the original quantum Zakharov system will be reported in a future work.   

\begin{figure}
\includegraphics{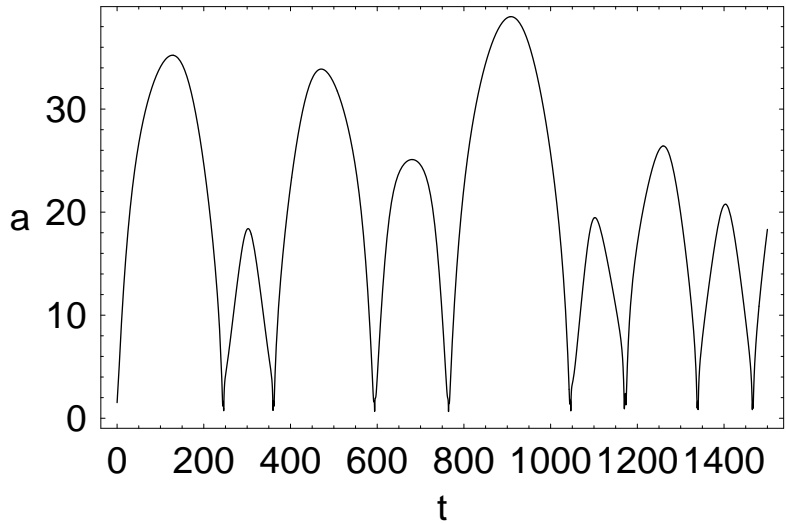}
\caption{\label{fig:epsart2} Simulation for the semiclassic system (\ref{e14})-(\ref{e16}) showing $a(t)$. Parameters, $M = N = 3$, $H = 0.3$. Initial condition, $(a_0, s_0, \dot{a}_0, \dot{s}_0) = (1.52, 1.36, 0, 0.62)$.}
\end{figure}

\begin{figure}
\includegraphics{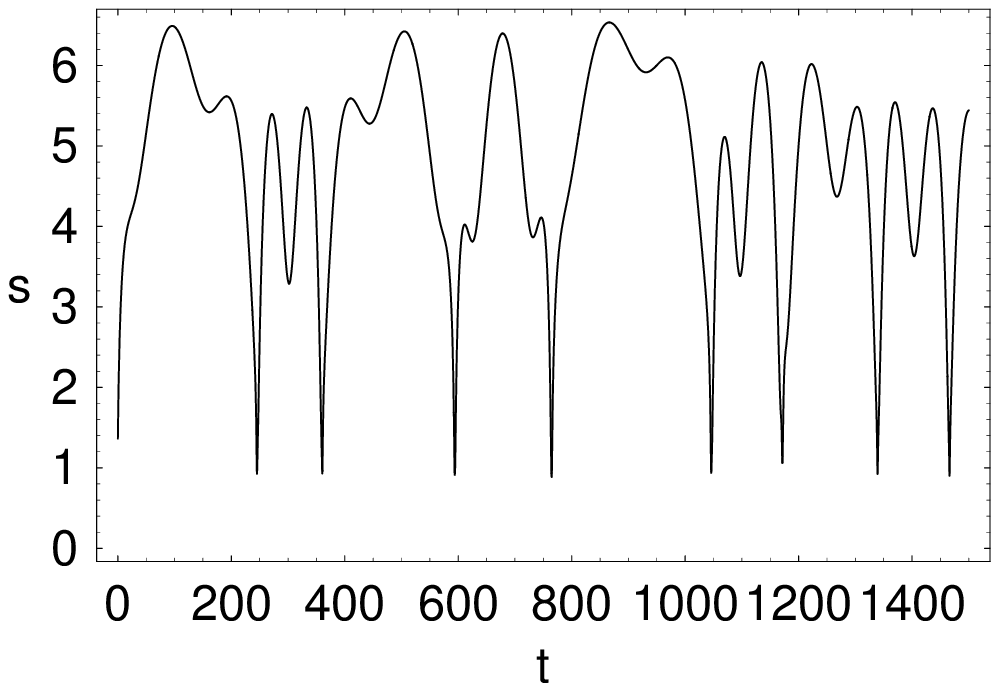}
\caption{\label{fig:epsart3} Simulation for the semiclassic system (\ref{e14})-(\ref{e16}) showing $s(t)$. Parameters, $M = N = 3$, $H = 0.3$. Initial condition, $(a_0, s_0, \dot{a}_0, \dot{s}_0) = (1.52, 1.36, 0, 0.62)$.}
\end{figure}

\begin{figure}
\includegraphics{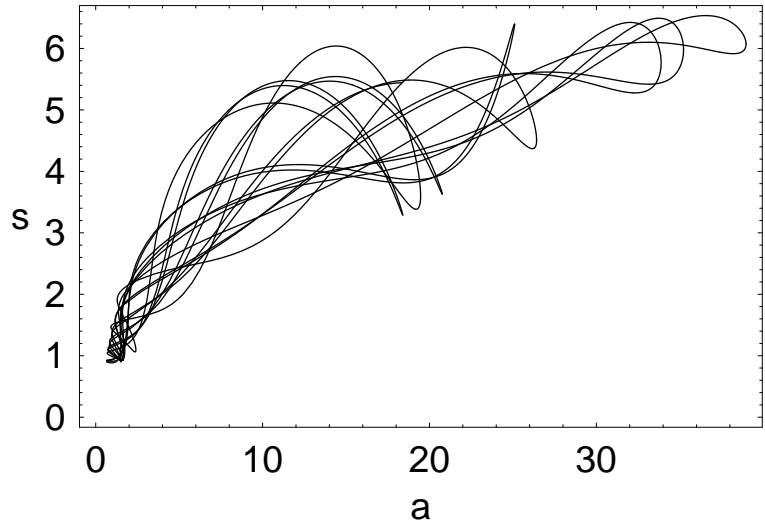}
\caption{\label{fig:epsart4} Trajectory for (\ref{e14})-(\ref{e16}) in configuration space. Parameters, $M = N = 3$, $H = 0.3$. Initial condition, $(a_0, s_0, \dot{a}_0, \dot{s}_0) = (1.52, 1.36, 0, 0.62)$.}
\end{figure}

\section{Non perturbative full system}
When $H$ is not a small parameter, one is not allowed to solve (\ref{e12}) retaining only the first-order quantum correction. Therefore, one is left with the full system (\ref{e12})-(\ref{e14}), for which some conclusions can be obtained. Even for the balanced case when $M = N$, it is not possible to get a closed-form solution for the fixed points, making difficult to derive general statements about linear stability. However, simulations can be made for different values of $H$, starting from a fixed point numerically calculated. For $M = N = 1, H = 5$, an equilibrium is found for $(\kappa, a, s) = (0, 9.15, 3.53)$. Figure 5 shows a typical trajectory starting at this initial condition, with $\dot{s}_0 = 0.2$. Under the same parameters but with a smaller initial velocity produces quasi-periodic motion, as shown in figure 6, where $\dot{s}_0 = 0.05$. Similar simulations shows that for increasing $H$ it becomes more difficult to get regular, quasi-periodic trajectories, pointing for instabilities of quantum nature. In addition, unbounded motion appears for smaller values of the initial velocity. 

\begin{figure}
\includegraphics{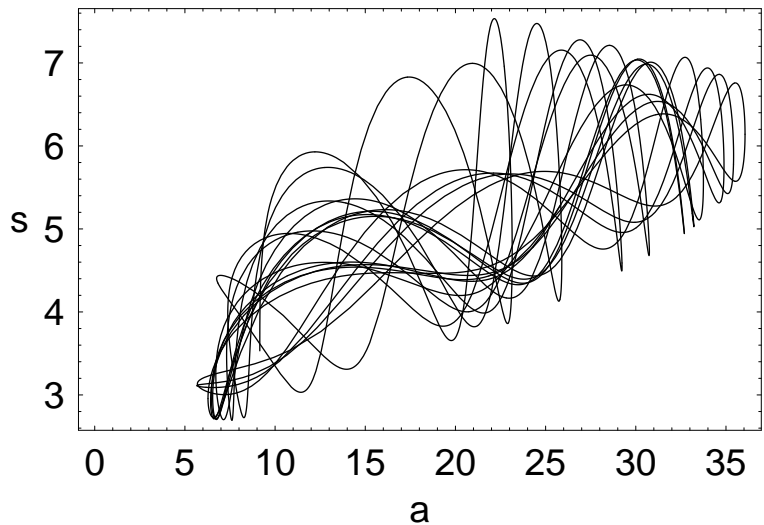}
\caption{\label{fig:epsart5} Simulation for the full dynamical system (\ref{e12})-(\ref{e14}) showing $a$ and $s$.   Parameters, $M = N = 1$, $H = 5$. Initial condition at $(\kappa, a, s, \dot{s}) = (0, 9.15, 3.53, 0.20)$.}
\end{figure}

\begin{figure}
\includegraphics{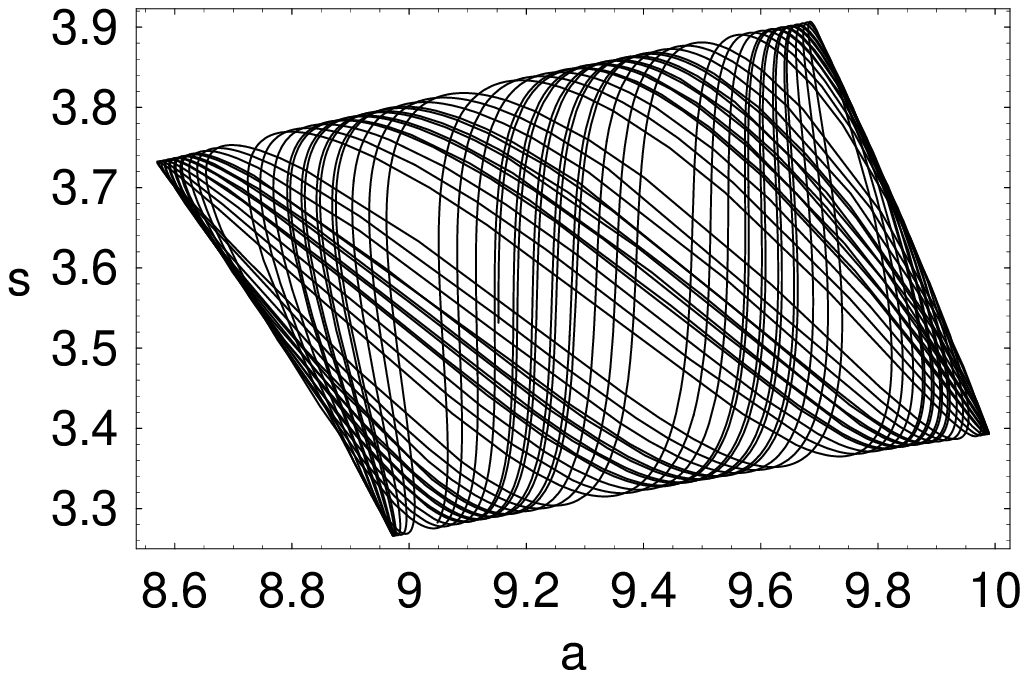}
\caption{\label{fig:epsart6} Simulation for the full dynamical system (\ref{e12})-(\ref{e14}) showing $a$ and $s$.   Parameters, $M = N = 1$, $H = 5$. Initial condition at $(\kappa, a, s, \dot{s}) = (0, 9.15, 3.53, 0.05)$.}
\end{figure}

Further results on the r\^ole of quantum effects can be obtained for the ultra-quantum case where we can neglect all terms at the right-hand sides of (\ref{e12})-(\ref{e14}) not containing $H^2$. In such case where $H^2$ is large enough, one get the system 
\begin{eqnarray}
\label{a}
\dot{a} &=& 12 H^2 \kappa (1 + 4 \kappa^2 a^4)/a \,,\\
\label{k}
\dot\kappa &=& 3 H^2 (1 - 16 \kappa^4 a^8)/a^6 \,,\\
\label{b}
\ddot{s} &=& 3 H^2 /s^7 \,.
\end{eqnarray}
Equations (\ref{a}-\ref{k}) can be solved yielding
\begin{equation}
\label{u}
a^2 = {a_{0}^{2}}  + \frac{36 H^4 (t - t_{0})^2}{{a_{0}^{6}} } \,, \quad \kappa^2 = \frac{9 {a_{0}^{4}}  H^4 (t - t_{0})^2}{({a_{0}^{8}} + 36 H^4 (t - t_{0})^2)^2} \,,
\end{equation}
where $a_{0}$ and $t_0$ are numerical constants. From (\ref{u}) the conclusion is that the width $a$ of the envelope electric field tends to increase without bound in the ultra-quantum limit, while the chirp function $\kappa$ approaches zero. Similarly, inspection of (\ref{b}) shows that the width $b = s^2$ of the density fluctuation increases without bound, since $\ddot{s} = - \partial V/\partial{s}$ for a pseudo-potential $V = H^{2}/2s^6$ having no bound-states. The results at the end of this section comes from the rough procedure of disregarding all terms not containing $H^2$ in the right-hand sides of  
(\ref{e12})-(\ref{e14}). They are consistent with the previous simulations, showing the destabilizing r\^ole of quantum effects.

\section{Conclusion}
The quantum Zakharov system was analyzed through a time-dependent Gaussian trial function method for an associated Lagrangian formalism. This extends to the quantum plasmas realm the results obtained for the classical Zakharov system by a similar approach \cite{Malomed}. In contrast to the classical case, complicated trajectories and instabilities can be found even for the balanced case of equal values of the invariants $M$ and $N$, corresponding to the low-frequency part of the mass and the number of high frequency quanta, respectively. Quantum effects plays a destabilizing r\^ole, yielding the ultimate decaying of the Langmuir soliton whose properties are simulated by the time-dependent Gaussian ansatz. This is a signature of quantum effects (wave-packet spreading, tunneling), making more difficult the existence of coherent, localized solutions in quantum plasmas. Direct comparison between the variational solutions of this work and numerical simulations of the original Zakharov system remains an open question.  

\noindent
{\bf Acknowledgments}

\noindent
We thanks the Brazilian agency Conselho Nacional de
Desenvolvimento Cien\-t\'{\i}\-fi\-co e Tec\-no\-l\'o\-gi\-co (CNPq) for financial support and Dra. Rejane Oliveski for aid with the figures.


\begin{thebibliography}{99}
\bibitem{Markowich} P. A. Markowich, C. A. Ringhofer and C. Schmeiser,   {\it Semiconductor Equations} (Springer, Vienna, 1990).
\bibitem{Rascol} G. Rascol, H. Bachau, V. T. Tikhonchuk, H. J. Kull and T. Ristow, Phys. Plasmas {\bf 13}, 103108 (2006).
\bibitem{Opher} M. Opher, L. O. Silva, D. E. Dauger {\it et al.},  Phys. Plasmas {\bf 8}, 2454 (2001).
\bibitem{H} F. Haas, L. G. Garcia, J. Goedert and G. Manfredi,  Phys. Plasmas {\bf 10}, 3858 (2003).
\bibitem{HH} F. Haas, Phys. Plasmas {\bf 12}, 062117 (2005).	
\bibitem{Haas} F. Haas, G. Manfredi and M. Feix, Phys. Rev. E  
{\bf 62}, 2763 (2000). 
\bibitem{Manfredi2} G. Manfredi and F. Haas, Phys. Rev. B {\bf 64},   075316 (2001).
\bibitem{Haas2} F. Haas, G. Manfredi and J. Goedert, Phys. Rev. E {\bf 64}, 26413 (2001). 
\bibitem{Shukla2} P. K. Shukla and L. Stenflo, New J. Phys. {\bf 8}, 111 (2006).
\bibitem{Stenflo} L. Stenflo, P. K. Shukla and M. Marklund, Europhys. Lett. {\bf 74}, 844 (2006).
\bibitem{Shukla5} P. K. Shukla and L. Stenflo, Phys. Lett. A {\bf 355}, 378 (2006).
\bibitem{Shukla6} P. K. Shukla and L. Stenflo,  Phys. Plasmas {\bf 13}, 044505 (2006).
\bibitem{Shukla7} P. K. Shukla and S. Ali, Phys. Plasmas {\bf 12}, 114502 (2005).
\bibitem{Ali1} S. Ali and P. K. Shukla, Phys. Plasmas {\bf 13}, 022313 (2006).
\bibitem{Misra} A. P. Misra and A. R. Chowdhury, Phys. Plasmas {\bf 13}, 072305 (2006).
\bibitem{mm} M. Marklund and G. Brodin, e-print physics/0612062. 
\bibitem{mmm} G. Brodin and M. Marklund, e-print physics/0612243.
\bibitem{Manfred} G. Manfredi, Fields Inst. Commun. {\bf 46}, 263 (2005).
\bibitem{Garcia} L. G. Garcia, F. Haas, J. Goedert and L. P. L. Oliveira, Phys. Plasmas {\bf 12}, 012302 (2005).
\bibitem{Zakharov} V. E. Zakharov, Sov. Phys. JETP {\bf 35}, 908 (1972).
\bibitem{Marklund} M. Marklund, Phys. Plasmas {\bf 12}, 082110 (2005).
\bibitem{Jovanovic} D. Jovanovic and R. Fedele, Phys. Lett. A, article in press (2007). 
\bibitem{Shukla} P. K. Shukla and B. Eliasson, Phys. Rev. Lett. {\bf 96}, 245001 (2006).
\bibitem{Yang} Q. Yang, C. Dai, Y. Wang and J. Zhang, J. Phys. Soc. Jpn. {\bf 74}, 2492 (2005).
\bibitem{Thornhill} S. G. Thornhill and D. ter Haar, Phys. Rep. {\bf 43}, 43 (1978).
\bibitem{Malomed} B. Malomed, D. Anderson, M. Lisak, M. L. Quiroga-Teixeiro and L. Stenflo, Phys. Rev. E {\bf 55}, 962 (1997). 
\bibitem{Anderson} D. Anderson, Phys. Rev. A {\bf 27}, 3135 (1983). 
\bibitem{bec} F. Haas, Phys. Rev. A {\bf 65}, 33603 (2002).
\bibitem{Sharma} R. P. Sharma, K. Batra and S. S. Das, Phys. Plasmas  {\bf 12}, 092303 (2005). 
\end{thebibliography}
\end{document}